# Cooling in QCD Spectroscopy

Robert G. Edwards[*], Khalil M. Bitar, Urs M. Heller, and A. D. Kennedy

SCRI, Florida State University, Tallahassee, FL 32306-4052, USA

Internet: {kmb,edwards,heller,adk}@scri.fsu.edu

We test the cooling algorithm with gluonic and staggered hadronic spectroscopy on $SU(3)$ gauge field configurations generated with two flavors of staggered dynamical fermions. We find cooling is not reliable as the basis for improved hadronic operators. We also find that performing cooling sweeps to reveal more clearly the topological properties of the gauge fields eliminates the spin structure of the hadron spectrum.

## 1. Introduction

Motivated by Shuryak et. al. [1] who have argued that instantons are the driving mechanism for particle mass generation in QCD, Chu, et. al. [2] used the cooling algorithm ($\beta = \infty$ Metropolis) on quenched $SU(3)$ configurations and measured hadronic spatial correlation functions and various gluonic quantities. They performed 0, 25 and 50 cooling sweeps. They claim to find reasonable agreement with the work of Shuryak et. al. for cooled and uncooled results, and argue in support of the dominant role of instantons. However, Trottier and Woloshyn [3] calculated hadronic spectroscopy on cooled $SU(2)$ configurations (which converge to a global minimum far more quickly) and large $m_q a$. They find the mesonic spectrum becomes degenerate after a few cooling sweeps.

Why use cooling? Operators that measure topological properties are plagued by ultraviolet (short distance) effects. Cooling eliminates these effects and exposes the metastable states of instantons, but the method has bias! Infinite cooling recovers a free field configuration. Cooling, or rather smearing, in dimensions orthogonal to time is valid for glueball states, but not for hadronic states. Teper [4] has argued that cooling is not uniform and short distance modes relax more quickly than long distance modes (which suffer from critical slowing down) to a global minimum. Therefore, we might expect to see short distance measurements converging to free field values faster than long distance measurements.

In this paper we will view the cooled gauge fields as composite fields (defined through the cooling algorithm!). We will test the hypothesis that hadronic and gluonic operators defined on these composite fields are "improved" operators: does the statistical improvement from (a few) cooling sweeps outway the inherent bias while at the same time exposing the long distance physics, namely instantons? Detailed results will be presented elsewhere [5].

For our tests, we used 200 of the $\beta = 5.6$, $m_q a = 0.01$ & $0.025$ 2-flavor staggered fermion gauge field configurations at SCRI generated by the HEMCGC collaboration. We performed extreme amounts of cooling sweeps, namely 0, 5, 25, 50, 100 and 500, to study its effects. For spectroscopy, we used (four flavor) staggered fermions with valence quark mass equal to the sea quark mass. We have also measured the eigenvalues of the Dirac operator (on selected configurations), $\langle \bar{\psi}\psi \rangle$, glueballs and Polyakov loops, and the heavy quark potential. All calculations were performed on the CM-2 at SCRI.

## 2. Simulations

In figures 1 and 2 we show the heavy quark potential $V(R)a$ calculated at time slices $T = 4, 5, 8, 10, 12$ for 5 and 25 cooling sweeps. We determine the lattice spacing by the force method of Sommer and compare to Ref. [6]. We find the string tension and lattice spacing are unchanged within errors between 0 and 5 cooling sweeps. However, after 25 cooling sweeps we see a clear separation of the potential from $T = 4$ and 5 but unchanged at $T >= 8$ except for an overall scale change – consistent with cooling altering short distance modes faster than long distance modes (non-uniform minimization).

---

[*]Speaker at the conference

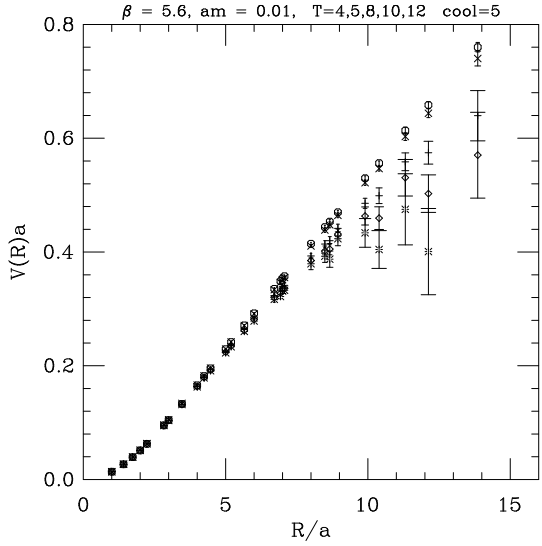

Figure 1. Heavy quark potential $V(R)a$ after 5 cooling sweeps. The potential is shown at $T = 4$ (○), $T = 5$ (×), $T = 8$ (+), $T = 10$ (◇), $T = 12$ (∗).

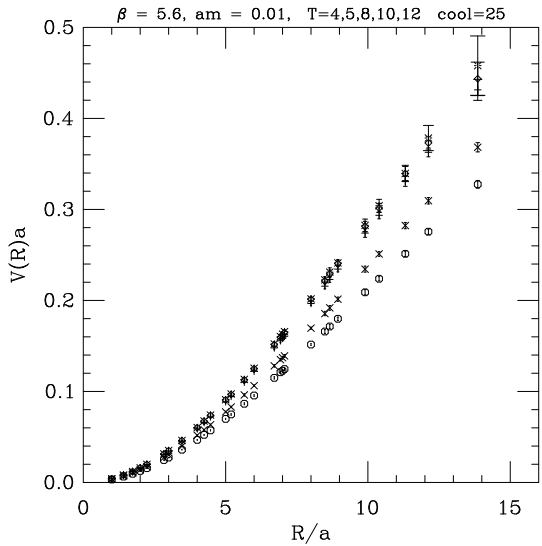

Figure 2. Heavy quark potential $V(R)a$ after 25 cooling sweeps. Symbols are the same as in Fig. 1. Note the separation of $V(R)a$ at smaller $T$.

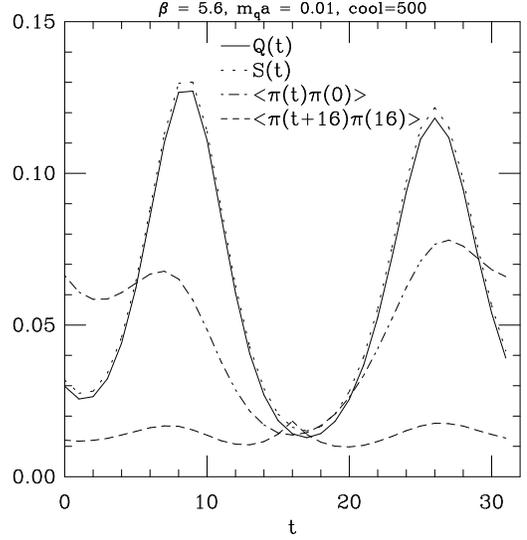

Figure 3. Time profile of $Q(t)$, $S(t)$, and the pion propagator $\langle \pi(t)\pi(0)\rangle$ and $\langle \pi(t+16)\pi(16)\rangle$.

| Cool | $a(r_0)$ | $m_\pi a$ | $m_\rho a$ | $m_N a$ |
|---|---|---|---|---|
| 0 | 0.096(4) | 0.269(2) | 0.516( 5) | 0.722(28) |
| 5 | 0.094(5) | 0.166(4) | 0.422(31) | 0.577(80) |
| 25 | 0.077(4) | 0.168(5) | 0.410(26) | 0.477( 6) |
| 50 | 0.067(3) | 0.177(8) | 0.369(15) | ——— |
| 100 | 0.057(3) | 0.194(8) | 0.330( 9) | ——— |
| 500 | ——— | 0.247(7) | 0.273(15) | ——— |

Table 1
Hadron masses at $m_q a = 0.01$. The confidence levels are not shown but are greater than 68%.

| Cool | $a(r_0)$ | $m_\pi a$ | $m_\rho a$ | $m_N a$ |
|---|---|---|---|---|
| 0 | 0.104(2) | 0.419(1) | 0.631( 6) | 0.968( 3) |
| 5 | 0.102(4) | 0.235(2) | 0.464(11) | 0.685( 2) |
| 25 | 0.082(4) | 0.216(2) | 0.426( 9) | 0.534( 6) |
| 50 | 0.074(3) | 0.209(3) | 0.393( 7) | 0.580(15) |
| 100 | 0.061(3) | 0.206(4) | 0.348( 7) | 0.568(24) |
| 500 | ——— | 0.233(7) | 0.267(19) | 0.493(15) |

Table 2
Hadron masses at $m_q a = 0.025$.

We also observed this phenomena in our measurements of the $0^{++}$, $1^{+-}$, $2^{++}$ glueballs masses using different levels of $3d$ smearing in addition to cooling. These operators are variational bound states, and if we had not smeared in the time direction we would expect to see the glueball effective mass bound decrease in time. However, at fixed cooling sweeps, we find the glueball effective mass increase in time, and at fixed time the effective mass decreases under cooling consistent again with non-uniform minimization. Comparing to Ref. [8], at $t = 2$, $m_{\text{eff}}(0^{++}) = 0.82(5), 0.46(2), 0.29(2), 0.24(2), 0.20(1), 0.118(8)$. Only at 500 cooling sweeps can we reliably extract $m(0^{++}) = 0.23(4)[0.40/12]$.

We calculated $\langle \bar{\psi}\psi \rangle$ using one Gaussian field. We found $\langle \bar{\psi}\psi \rangle = 0.1122(5), 0.0202(4), 0.0283(6), 0.0278(7), 0.0265(9), 0.0249(8)$ for $m_q a = 0.01$, and $\langle \bar{\psi}\psi \rangle = 0.2140(3), 0.0589(2), 0.0541(2), 0.0523(2), 0.0508(2), 0.0491(2)$ for $m_q a = 0.025$ as a function of cooling. We see $\langle \bar{\psi}\psi \rangle$ is decreasing but not as fast as in $SU(2)$ [3] – we see no evidence that the instanton induced zero modes for staggered fermions are not becoming delocalized as required for chiral symmetry breaking [7].

In Fig. 3 we show a time profile of the topological charge density and energy density summed over 3-space, along with the wall source point sink zero momentum pion propagator with wall source at $t = 0$ and 16 on a single configuration at 500 cooling sweeps. We see the charge and energy density (normalized in units of a single instanton) significantly overlap and correspond to large bumps in the pion propagator. Therefore, to enhance statistics and restoration of translation invariance we measured our spectroscopy with 16 time-slice sources. We show the results for $m_q a = 0.01$ and $0.025$ in Tables 1 and 2. We see the mesons for both $m_q a$ are becoming degenerate indicating that cooling has eliminated spin structure. In addition, flavor symmetry is restored after 5 cooling sweeps. The nucleon is consistently about 3/2 times larger than the rho.

## 3. Conclusions

We find cooling is *not* reliable as an improved hadronic operator. A few cooling sweeps leaves long distance gluonic physics intact (as is well known) but has dramatic effects on the staggered hadron spectrum. As in the results of ref. [3], we find that performing more cooling sweeps to reveal more clearly the topological properties of the gauge fields eliminates the spin structure of the hadron spectrum. As a function of cooling we find the pion mass makes a precipitous drop (indicating heavy mixing with short range fluctuations) then slowly rises. All other particle masses decrease uniformly and the overall spectrum becomes degenerate. We did not perform enough cooling sweeps for the spectrum to become free field like.

It is not clear what has been revealed about the instanton content of the QCD vacuum. Instantons definitely affect quark propagators as can be seen from the eigenvectors of the dirac operator or the profiles of the pion propagator (Fig. 3). This is not too surprising since very little energy content is left. However, cooling is definitely eliminating staggered fermion spin structure which is undoubtedly related to the large energy remaining after a few cooling sweeps. Without an a priori cutoff, the utility of the cooling algorithm is questionable.